\documentclass[twocolumn,aps,floats,floatfix,nofootinbib,prd,superscriptaddress,tightenlines]{revtex4}
\usepackage{graphicx}
\usepackage{color}

\begin{document}

\title{Using jet mass to discover vector quarks at the LHC}
\author{Witold Skiba}
\affiliation{Department of Physics, Yale University, New Haven, CT 06520}
\author{David Tucker-Smith}
\affiliation{Department of Physics, Williams College, Williamstown, MA 01267}

\begin{abstract}
We illustrate the utility of jet mass distributions as probes of  new physics
at the LHC, focusing on a heavy vector-quark doublet that mixes with the top as a concrete example.
For 1 TeV vector-quark masses, we find that signals with greater than $5 \sigma$ significance can
be achieved after 100 fb$^{-1}$. More generally, jet mass distributions have the potential to  provide
signals for heavy states that produce highly boosted weak gauge bosons and/or top quarks.

\end{abstract}

\maketitle
\section{introduction}
Various extensions of the Standard Model include
additional quarks in vector representations of the standard model gauge 
group. Examples include little Higgs theories  \cite{lh,littlest,SU6Sp6}, topcolor models 
\cite{tc}, and extra dimensional theories with bulk fermions.  Vector-like matter is also 
invoked in some supersymmetric models, for instance as the messengers
of supersymmetry breaking in gauge-mediated models \cite{gm}.
If  light enough, vector-like quarks  would be produced copiously at 
the LHC, and the details of how they might be discovered would depend on how, 
or whether, they decay.

In this paper we concentrate on heavy quarks 
that decay into gauge bosons and top quarks.
In particular, we will be interested in the case where the quarks 
are so heavy that the $W$'s, $Z$'s, and tops which they produce in turn 
yield highly collimated decay products that  cannot typically  be 
resolved into separate jets.
The invariant masses of individual jets then become potentially useful quantities
to study when attempting to pick out signals.  
Jet mass was originally used as a probe of QCD \cite{Clavelli}, and has since
been  shown to be useful in studies of elastic WW scattering at high
energy~\cite{BCF}.  
Here we employ jet mass to provide signals for  vector-quarks, and
show that its usefulness persists after detector effects are taken into account.
It is clear that jet mass could also be helpful in
other collider studies with similar kinematic properties.

One well-studied scenario with vector-quarks features electroweak 
singlets $T+{\overline T}$, with hypercharge $\pm 2/3$.
This is the extra fermion content in the littlest Higgs model  \cite{littlest}.
If  $T$ mixes with the the top quark, then in the regime where the 
vector-quark mass is much larger
than the top mass, the branching ratios for the decays of $T$ are 
approximately $Br(T \rightarrow b W^+)\simeq 2 Br(T \rightarrow tZ) 
\simeq 2 Br(T \rightarrow th) \simeq 1/2$.  Moreover, 
provided that the mixing is large enough, the $T$ quarks can be 
produced singly by $t$-channel exchange of a $W$ boson, with a $b$ 
quark in the initial state \cite{littlestpheno}, the importance of the this being 
that the cross section for single production falls off less 
dramatically with increasing $T$ mass than the cross section for 
$T{\overline T}$ pair production.  In \cite{Azuelos:2004dm}, it was estimated that 
for order-one mixing, the discovery reach
  after an integrated luminosity of 300$^{-1}$ fb at the LHC is $\simeq 
2$--2.5 TeV.    On the other hand, if the mixing is small, or if the 
$T$ quark is light enough,  then QCD pair-production dominates.  This 
case was studied in \cite{Aguilar-Saavedra:2005pv}, where it was estimated that the 
discovery reach after 300$^{-1}$ fb  at the LHC is  $\simeq 1.1$ TeV.

Here we study the case where the vector-quarks are instead 
electroweak doublets $Q+{\overline Q}$, with hypercharge $\pm 2/3$.   
Electroweak doublet vector-quarks appear, for instance, in the little 
Higgs model of  \cite{SU6Sp6}, and the topcolor model of \cite{Popovic:2001cj}.
  We will argue below that it is reasonable to imagine that the decays of 
$T$ and $B$, the  upper and lower components of $Q$,
   are induced by the mixing of $T$ with the top quark.  In this case, 
$B$ decays  to $t W^-$, just as if it were a fourth-generation down 
quark.
   The prospects for discovering such a particle at the LHC have been
  explored in \cite{tdr}.  The approach taken there is to search for 
$W$ candidates by looking at dijet invariant masses,
  finding top candidates by looking at the invariant masses of the the 
$W$ candidates and $b$-tagged jets,
  and finally, looking at the invariant mass distribution for the $W$ and 
top candidates.
  The peak in this distribution ends up being rather broad, so that even 
for a $B$ mass of 640 GeV
  and an integrated luminosity of 100$^{-1}$ fb, separation of signal 
from background looks challenging.

Our strategy, which we will apply to TeV-mass $B$'s, will be to find signals in jet
mass distributions.  We will see that these distributions feature
bumps around  $m_W$ and $m_t$,
  and that the invariant mass distribution for the candidate top and $W$ 
jets is peaked near the vector-quark mass.

In the next section we outline the model and list  our assumptions 
about its free parameters. The main analysis, requiring at least one high $p_T$ lepton, 
is presented in  Sec.~III. There we list the relevant backgrounds and propose cuts 
that give a convincing signal. A dilepton analysis is presented in Sec.~IV.

\section{the model}

Our vector quark doublet $Q+{\overline Q}$
has a mass term and also Yukawa couplings with the third generation 
quarks,
\begin{eqnarray}
{\mathcal L_{mass}} &  = & M Q{\overline Q}+
(\lambda_u Q u_3^c {\tilde h}+
  \lambda_u' q_3 u_3^c {\tilde h} + \nonumber\\
    & & \quad \quad \quad \quad \quad  \lambda_d Q d_3^c h+
     \lambda_d'  q_3 d_3^c h + {\rm h.c.}).
\end{eqnarray}
Electroweak symmetry breaking induces mixing of the heavy quarks with the third generation.
We neglect mixings of the vector quark with the lighter two  generations.  
The main motivation for doing so is to simplify the analysis,
but the quark mass hierarchies themselves make this a reasonable starting point.
 Moreover, the mixings with the lighter generations are more tightly constrained 
 experimentally.    Without loss of generality,
we have defined $Q$ to be the single linear combination of  $Y=1/6$ 
doublets that couples directly to ${\overline Q}$,
so there is no $q_3 {\overline Q}$ mass term.

If we adopt the reasonable assumption that $\lambda_u$ and $\lambda_u'$ 
are of comparable size, and so are $\lambda_d$ and $\lambda_d'$,  then in the heavy 
vector-quark regime that interests us, $M \gg m_{t}$, we have
\begin{equation}
   m_{t} \simeq \lambda_u' v, \quad m_{b} \simeq \lambda_d' v, \quad 
   m_{T}\simeq m_{B} \simeq M,
\end{equation}
where $m_T$ and $m_B$ are the masses of  $T$ and $B$, the up and down 
components of $Q$. The ratio of the couplings of $Q$ to the bottom
and top quarks is then  approximately
\begin{equation}
    (\lambda_d/\lambda _d')\times (\lambda_u'/\lambda _u)\times 
(m_b/m_t).
\end{equation}
We will assume that this product is sufficiently small that decays of 
$B$ and $T$ directly into bottom quarks can be neglected.   This 
assumption is motivated by the fact that $m_b$ is much smaller than 
$m_t$, and also by the fact that mixing with the bottom
is constrained by $Z$-pole data.  Mixing with the top is 
less constrained, as the electroweak couplings of the top quark
have not been measured precisely.   In this case, the equivalence
principle tells us that in the large  $M$ limit,
the branching ratios of $B$ and $T$ are
\begin{equation}
  Br(B \rightarrow t W-)  \simeq   100\%
\end{equation}
\vspace{-0.3in}
\begin{equation}
   Br(T \rightarrow t Z) \simeq  Br(T \rightarrow t h) \simeq 50\%
\end{equation}
Thus $B$ decays just as if it were a fourth-generation down quark, 
while $T$ has two possible final states, and all decays produce top 
quarks. 

Although electroweak symmetry lifts the degeneracy between $B$ 
and $T$, the splitting  $\delta$ is  only
\begin{equation}
   \delta \equiv m_T -m_B \simeq 15 \; {\rm GeV} \times \left({\lambda_u 
   \over \lambda_u'}\right)^2 \times \left({1\;{\rm TeV} \over M} \right),
\end{equation}
and the rate for the three-body decays $T \rightarrow B  \; f  {\overline f'}$ 
is proportional to $\delta^5/ v^4$.  Even neglecting 
phase space factors, the ratio of this rate to the two-body rates is 
proportional to $(m_t/M)^6$, and so we will neglect the three-body
decay entirely in  what follows.

\section{single lepton analysis}
We now explore the ability of the LHC to probe this model through final states
with at least one lepton.  For concreteness, we will fix $M$ to be 1 TeV.   Our discussion 
will focus on the production and decay of $B$ particles, as our method will be far more
sensitive to $B$'s than to $T$'s.  For our calculations we will  take the Higgs mass to be 120 GeV,
although our results are not very sensitive to its value.  

The heavy $B$ can be produced singly in association with a top quark,
with a cross section that depends on the heavy quark mass and on the amount of mixing.  
For large $M$, the cross section is roughly proportional to $(\lambda_u v)^2/M^2$. 
Setting $\lambda_u v = m_t$ and $M=1$ TeV, we find using Madgraph~\cite{Maltoni:2002qb} 
that the leading-order cross section for single $B$/${\overline B}$ production is  $14$ fb.  
By comparison, the  pair production cross section, which depends only on the heavy 
quark mass, is $60$ fb at NLO with gluon resummation \cite{Bonciani:1998vc}.  
In what follows, we will focus on the pair production process exclusively. 
It is possible that single production may allow discovery of heavier $B$'s if the mixing angle
is large enough; on the other hand, if the mixing angle is much smaller than $m_t/M$, 
single production is not likely to be of any help at all.

In our analysis, we use MadGraph/MadEvent \cite{Maltoni:2002qb} to generate signal
events at the parton level, taking the renormalization and factorization scales to be twice
the heavy quark mass, and rescaling the cross section to agree with the NLO result.  
These are passed to Pythia 6.325 \cite{Sjostrand:2003wg} to simulate initial and final-state
radiation, multiple interactions, and hadronization.  We use Alpgen 2.06 \cite{Mangano:2002ea} 
and Pythia 6.325 to generate background events, and to obtain inclusive event samples 
we apply the  MLM prescription for jet-parton matching  \cite{MLM}  as implemented in  Alpgen 2.11. 
The one exception is that jet-parton matching is not performed for single-top processes.   
For all calculations we use the PDF set CTEQ5L \cite{Lai:1999wy}. 

We pass  showered events to the PGS-4 detector simulator \cite{PGS}.  
Since we are interested in jet mass, for our purposes the most crucial settings of the detector simulator 
are the energy resolution of the hadronic calorimeter, which we take to be $\Delta E/E = 0.8/\sqrt{E/{\rm GeV}}$, 
and the granularity  of the calorimeter, which we take to be $\Delta \phi \times \Delta \eta = 0.1 \times  0.1$.  
The PGS detector simulator uses the $k_T$ algorithm for jet clustering \cite{Catani:1991hj}.   
We adopt reference cone size  $\Delta R_{cone} = 0.5$, and set the threshold transverse energy 
for a cell to be included in the clustering at 5 GeV.  This large threshold leads to a slight underestimation 
of jet energies, but turns out to be helpful for reducing the degradation of the signal from multiple interactions. 
We use the heavy-flavor tagging efficiencies included in the PGS code, which are based
on the results of a vertexing algorithm applied to CDF calibration data. 
Finally, the lepton isolation criteria are as follows: the total $p_T$ of tracks within
$\Delta R =0.4$ of the lepton is required to be less than 5 GeV, and
the total $E_T$ in the $3 \times 3$ calorimeter grid with the lepton's cell at the center 
(excluding the $E_T$ of that central cell) is required to be less than $0.1$ or $0.1125$
 times the $E_T$ in the central cell, for electrons and muons, respectively. 

To suppress backgrounds, we adopt the following cuts:
\begin{itemize}
\item ${p_T \! \! \! \! \! \!\slash} \;+ \sum p_T > 1800\;$ GeV, where the sum is over all photons, leptons,
 and jets having  $p_T > 20 \;$ GeV and $|\eta| < 2.5$

\item at least one lepton ($e$, $\mu$) with $p_T > 100 \;$ GeV and $|\eta| < 2.5$

\item ${p_T \! \! \! \! \! \!\slash} \;> 100 \;$ GeV

\item at least one $b$-tagged jet with $p_T > 20$

\item $\Delta R_{lj}>1.0$, where  $\Delta R_{lj}$ is the separation between the hardest lepton and the closest
jet having $p_T>$ 20 GeV.  This cut is useful for reducing the $t {\overline t}$ background, 
in which the leptons produced by the highly boosted tops are typically quite close to the $b$ quarks.  

\item $S_T>0.1$, where $S_T$ is the transverse sphericity.  Given the $2\times 2$ tensor 
$S_{ij}=\sum_{\alpha} p^{\alpha}_i p^{\alpha}_j$, where $i$ and $j$ label the two directions 
perpendicular to the beam, and $\alpha$ labels the jets, leptons, and photons having 
$p_T > 20 \;$ GeV and $|\eta| < 2.5$, along with the missing $p_T$, we define $S_T$ 
as twice the smaller eigenvalue divided by the trace (so $0\leq S_T \leq 1$ is always satisfied).
\end{itemize}
After these cuts, the dominant backgrounds are $W$+ jets, $t {\overline t} $, and  $W {b \overline b}$, 
followed by $WW$ and $tW$. We will assume that with these cuts we can neglect the background
from QCD with a fake lepton, although to seriously address this background source would
presumably require a study of fake rates using LHC data.

To obtain large enough background samples, we impose the generator-level (pre- showering/hadronization) 
cut on all  background processes except for single-top production: 
\begin{equation}
{p_T \! \! \! \! \! \!\slash} \;+ \sum p_T > 1500\; {\rm GeV},
\end{equation}
where the sum is over the final-state partons ({\em e.g.} $t$, ${\overline t}$, and any extra light
quarks or gluons  for the $t {\overline t}$+ jets sample).  We estimate how much of the background 
we lose by generating event samples below this cut, seeing what fraction pass the 
${p_T \! \! \! \! \! \!\slash} \;+ \sum p_T > 1800\;$ GeV  cut  after detector simulation, 
and assuming that the rest of the cuts have roughly the same effect on the remaining 
events as for the sample that passed the generator-level cut.  In this way we  estimate that 
well under 10\% of the background is neglected due to this generator-level cut.     

For the single-top final states $tW$, $tq$, $tb$, and $tbW$, multiple extra jets are not included at 
the matrix-element level, and we find that it is necessary to loosen this cut.  After relaxing it to 
\begin{equation}
{p_T \! \! \! \! \! \!\slash} \;+ \sum p_T > 1000\; {\rm GeV}
\end{equation}
for these processes, we again estimate that well under 10\% of the background is neglected.  

For all processes for which jet-parton matching is performed, we impose the following generator-level cuts 
on the extra light jets: $p_T > 20$ GeV, $|\eta| < 2.5$, and a jet-jet separation $\Delta R_{jj} > 0.7$ 
(cuts of this nature are part of the jet-parton matching program).  For $W b {\overline b}$, the minimum
separation between extra light-jets and bottom quarks is also set to 0.7.  
The minimum cluster $E_T$, rapidity range, and $\Delta R$ used 
for the jet-parton matching are then set to their Alpgen defaults of 25 GeV,  $|\eta| < 2.5$, and 0.7, respectively.  
For $t W$ jet-parton matching is not performed, and the extra jet in the $tW$+jet sample is required to have $p_T > 20$~GeV.    

For each background process, the factorization and renormalization scale $Q$  is set to its Alpgen
default, $Q=\sum m_B^2+ \sum m_T^2$, where the first term is the sum of the masses
of any final-state gauge bosons, the second sum is over all final-state
partons excluding the gauge bosons or their decay products, 
and where $m_T^2=p_T^2 + m^2$,  with $m$ being the mass of the parton.

In table \ref{table:first} we list the numbers of events generated for 
the various  backgrounds,  the cross sections and corresponding integrated luminosities,
and the numbers of events that pass all cuts.  The cross sections
are obtained by multiplying the Monte Carlo results by NLO $K$ factors.   The
factors we use are just rough estimates of the
NLO effects, but neglecting them would certainly underestimate the 
backgrounds.  We take $K=1.5$ for $t\bar{t}$~\cite{ttbarKfactor}, $K=1.25$ for
$W/Z+{\rm light\ jets}$~\cite{WZjetsKfactor},  $K=2$ for  $Wb\bar{b}$~\cite{WZjetsKfactor},
$K=1.1$ for  $WW$, $W,Z$, and $ZZ$~\cite{2VKfactors}, $K=1.1$ for $tq$~\cite{tqarkKfactors},
$K=1.5$ for $tb$~\cite{tqarkKfactors},  $K=1.2$ for $tW$~\cite{tWKfactor}, and  $K=1.5$ for $tbW$, 
although we are not aware of NLO results for this last final state.
In  generating these samples, the  states $WW$, $tbW$, and $tW$ are decayed inclusively, the
 states $t {\overline t}$, $ZZ$, and $WZ$, are  required to produce at least one lepton in their decays,  and
 leptonic decays are required for the gauge boson or top quark in the 
 $W$+jets, $Z$+jets, $Wb{\overline b}$+ jets, $tq$, and $tb$ samples.  
 
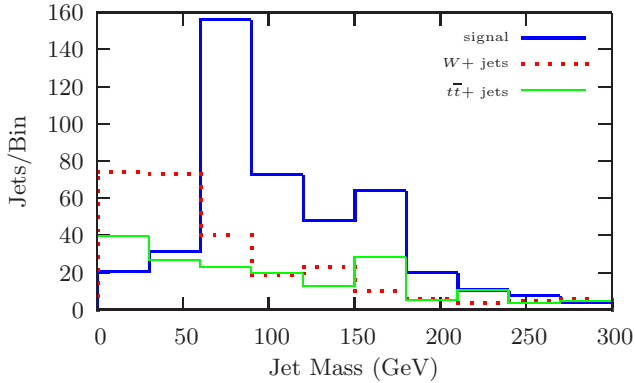
\begin{figure}[htbp] 
   \centering
\setlength{\unitlength}{0.240900pt}
\ifx\plotpoint\undefined\newsavebox{\plotpoint}\fi
\begin{picture}(1050,630)(0,0)
\sbox{\plotpoint}{\rule[-0.200pt]{0.400pt}{0.400pt}}%
\put(181.0,123.0){\rule[-0.200pt]{4.818pt}{0.400pt}}
\put(161,123){\makebox(0,0)[r]{ 0}}
\put(969.0,123.0){\rule[-0.200pt]{4.818pt}{0.400pt}}
\put(181.0,181.0){\rule[-0.200pt]{4.818pt}{0.400pt}}
\put(161,181){\makebox(0,0)[r]{ 20}}
\put(969.0,181.0){\rule[-0.200pt]{4.818pt}{0.400pt}}
\put(181.0,240.0){\rule[-0.200pt]{4.818pt}{0.400pt}}
\put(161,240){\makebox(0,0)[r]{ 40}}
\put(969.0,240.0){\rule[-0.200pt]{4.818pt}{0.400pt}}
\put(181.0,298.0){\rule[-0.200pt]{4.818pt}{0.400pt}}
\put(161,298){\makebox(0,0)[r]{ 60}}
\put(969.0,298.0){\rule[-0.200pt]{4.818pt}{0.400pt}}
\put(181.0,357.0){\rule[-0.200pt]{4.818pt}{0.400pt}}
\put(161,357){\makebox(0,0)[r]{ 80}}
\put(969.0,357.0){\rule[-0.200pt]{4.818pt}{0.400pt}}
\put(181.0,415.0){\rule[-0.200pt]{4.818pt}{0.400pt}}
\put(161,415){\makebox(0,0)[r]{ 100}}
\put(969.0,415.0){\rule[-0.200pt]{4.818pt}{0.400pt}}
\put(181.0,473.0){\rule[-0.200pt]{4.818pt}{0.400pt}}
\put(161,473){\makebox(0,0)[r]{ 120}}
\put(969.0,473.0){\rule[-0.200pt]{4.818pt}{0.400pt}}
\put(181.0,532.0){\rule[-0.200pt]{4.818pt}{0.400pt}}
\put(161,532){\makebox(0,0)[r]{ 140}}
\put(969.0,532.0){\rule[-0.200pt]{4.818pt}{0.400pt}}
\put(181.0,590.0){\rule[-0.200pt]{4.818pt}{0.400pt}}
\put(161,590){\makebox(0,0)[r]{ 160}}
\put(969.0,590.0){\rule[-0.200pt]{4.818pt}{0.400pt}}
\put(181.0,123.0){\rule[-0.200pt]{0.400pt}{4.818pt}}
\put(181,82){\makebox(0,0){ 0}}
\put(181.0,570.0){\rule[-0.200pt]{0.400pt}{4.818pt}}
\put(316.0,123.0){\rule[-0.200pt]{0.400pt}{4.818pt}}
\put(316,82){\makebox(0,0){ 50}}
\put(316.0,570.0){\rule[-0.200pt]{0.400pt}{4.818pt}}
\put(450.0,123.0){\rule[-0.200pt]{0.400pt}{4.818pt}}
\put(450,82){\makebox(0,0){ 100}}
\put(450.0,570.0){\rule[-0.200pt]{0.400pt}{4.818pt}}
\put(585.0,123.0){\rule[-0.200pt]{0.400pt}{4.818pt}}
\put(585,82){\makebox(0,0){ 150}}
\put(585.0,570.0){\rule[-0.200pt]{0.400pt}{4.818pt}}
\put(720.0,123.0){\rule[-0.200pt]{0.400pt}{4.818pt}}
\put(720,82){\makebox(0,0){ 200}}
\put(720.0,570.0){\rule[-0.200pt]{0.400pt}{4.818pt}}
\put(854.0,123.0){\rule[-0.200pt]{0.400pt}{4.818pt}}
\put(854,82){\makebox(0,0){ 250}}
\put(854.0,570.0){\rule[-0.200pt]{0.400pt}{4.818pt}}
\put(989.0,123.0){\rule[-0.200pt]{0.400pt}{4.818pt}}
\put(989,82){\makebox(0,0){ 300}}
\put(989.0,570.0){\rule[-0.200pt]{0.400pt}{4.818pt}}
\put(181.0,123.0){\rule[-0.200pt]{194.647pt}{0.400pt}}
\put(989.0,123.0){\rule[-0.200pt]{0.400pt}{112.500pt}}
\put(181.0,590.0){\rule[-0.200pt]{194.647pt}{0.400pt}}
\put(181.0,123.0){\rule[-0.200pt]{0.400pt}{112.500pt}}
\put(60,356){\makebox(0,0){{\rotatebox{90}{Jets/Bin}}}}
\put(585,30){\makebox(0,0){Jet Mass (GeV)}}
\sbox{\plotpoint}{\rule[-0.400pt]{0.800pt}{0.800pt}}%
\sbox{\plotpoint}{\rule[-0.200pt]{0.400pt}{0.400pt}}%
\put(829,550){\makebox(0,0)[r]{{\tiny signal}}}
{\color{blue}\sbox{\plotpoint}{\rule[-0.400pt]{0.800pt}{0.800pt}}%
\put(849.0,550.0){\rule[-0.400pt]{24.090pt}{0.800pt}}
\put(181.0,123.0){\rule[-0.400pt]{0.800pt}{14.454pt}}
\put(181.0,183.0){\rule[-0.400pt]{19.513pt}{0.800pt}}
\put(262.0,183.0){\rule[-0.400pt]{0.800pt}{7.709pt}}
\put(262.0,215.0){\rule[-0.400pt]{19.513pt}{0.800pt}}
\put(343.0,215.0){\rule[-0.400pt]{0.800pt}{87.688pt}}
\put(343.0,579.0){\rule[-0.400pt]{19.272pt}{0.800pt}}
\put(423.0,336.0){\rule[-0.400pt]{0.800pt}{58.539pt}}
\put(423.0,336.0){\rule[-0.400pt]{19.513pt}{0.800pt}}
\put(504.0,263.0){\rule[-0.400pt]{0.800pt}{17.586pt}}
\put(504.0,263.0){\rule[-0.400pt]{19.513pt}{0.800pt}}
\put(585.0,263.0){\rule[-0.400pt]{0.800pt}{11.563pt}}
\put(585.0,311.0){\rule[-0.400pt]{19.513pt}{0.800pt}}
\put(666.0,182.0){\rule[-0.400pt]{0.800pt}{31.076pt}}
\put(666.0,182.0){\rule[-0.400pt]{19.513pt}{0.800pt}}
\put(747.0,154.0){\rule[-0.400pt]{0.800pt}{6.745pt}}
\put(747.0,154.0){\rule[-0.400pt]{19.272pt}{0.800pt}}
\put(827.0,145.0){\rule[-0.400pt]{0.800pt}{2.168pt}}
\put(827.0,145.0){\rule[-0.400pt]{19.513pt}{0.800pt}}
\put(908.0,135.0){\rule[-0.400pt]{0.800pt}{2.409pt}}
\put(908.0,135.0){\rule[-0.400pt]{19.513pt}{0.800pt}}
\put(989.0,123.0){\rule[-0.400pt]{0.800pt}{2.891pt}}}
\sbox{\plotpoint}{\rule[-0.500pt]{1.000pt}{1.000pt}}%
\sbox{\plotpoint}{\rule[-0.200pt]{0.400pt}{0.400pt}}%
\put(829,509){\makebox(0,0)[r]{{\tiny $W$+ jets}}}
{\color{red}\sbox{\plotpoint}{\rule[-0.500pt]{1.000pt}{1.000pt}}%
\multiput(849,509)(20.756,0.000){5}{\usebox{\plotpoint}}
\put(949,509){\usebox{\plotpoint}}
\multiput(181,123)(0.000,20.756){11}{\usebox{\plotpoint}}
\multiput(181,340)(20.756,0.000){4}{\usebox{\plotpoint}}
\multiput(262,337)(20.756,0.000){4}{\usebox{\plotpoint}}
\multiput(343,337)(0.000,-20.756){5}{\usebox{\plotpoint}}
\multiput(343,241)(20.756,0.000){3}{\usebox{\plotpoint}}
\multiput(423,241)(0.000,-20.756){3}{\usebox{\plotpoint}}
\multiput(423,177)(20.756,0.000){4}{\usebox{\plotpoint}}
\put(504.00,179.69){\usebox{\plotpoint}}
\multiput(504,190)(20.756,0.000){4}{\usebox{\plotpoint}}
\multiput(585,190)(0.000,-20.756){2}{\usebox{\plotpoint}}
\multiput(585,153)(20.756,0.000){4}{\usebox{\plotpoint}}
\multiput(666,140)(20.756,0.000){4}{\usebox{\plotpoint}}
\multiput(747,133)(20.756,0.000){4}{\usebox{\plotpoint}}
\multiput(827,137)(20.756,0.000){4}{\usebox{\plotpoint}}
\put(908.00,139.06){\usebox{\plotpoint}}
\multiput(908,140)(20.756,0.000){3}{\usebox{\plotpoint}}
\put(989.00,138.91){\usebox{\plotpoint}}
\put(989,123){\usebox{\plotpoint}}}
\sbox{\plotpoint}{\rule[-0.200pt]{0.400pt}{0.400pt}}%
\put(829,468){\makebox(0,0)[r]{{\tiny $t {\overline t}$+ jets}}}
{\color{green}\put(849.0,468.0){\rule[-0.200pt]{24.090pt}{0.400pt}}
\put(181.0,123.0){\rule[-0.200pt]{0.400pt}{27.944pt}}
\put(181.0,239.0){\rule[-0.200pt]{19.513pt}{0.400pt}}
\put(262.0,201.0){\rule[-0.200pt]{0.400pt}{9.154pt}}
\put(262.0,201.0){\rule[-0.200pt]{19.513pt}{0.400pt}}
\put(343.0,190.0){\rule[-0.200pt]{0.400pt}{2.650pt}}
\put(343.0,190.0){\rule[-0.200pt]{19.272pt}{0.400pt}}
\put(423.0,181.0){\rule[-0.200pt]{0.400pt}{2.168pt}}
\put(423.0,181.0){\rule[-0.200pt]{19.513pt}{0.400pt}}
\put(504.0,160.0){\rule[-0.200pt]{0.400pt}{5.059pt}}
\put(504.0,160.0){\rule[-0.200pt]{19.513pt}{0.400pt}}
\put(585.0,160.0){\rule[-0.200pt]{0.400pt}{11.081pt}}
\put(585.0,206.0){\rule[-0.200pt]{19.513pt}{0.400pt}}
\put(666.0,138.0){\rule[-0.200pt]{0.400pt}{16.381pt}}
\put(666.0,138.0){\rule[-0.200pt]{19.513pt}{0.400pt}}
\put(747.0,138.0){\rule[-0.200pt]{0.400pt}{3.854pt}}
\put(747.0,154.0){\rule[-0.200pt]{19.272pt}{0.400pt}}
\put(827.0,133.0){\rule[-0.200pt]{0.400pt}{5.059pt}}
\put(827.0,133.0){\rule[-0.200pt]{19.513pt}{0.400pt}}
\put(908.0,133.0){\rule[-0.200pt]{0.400pt}{0.723pt}}
\put(908.0,136.0){\rule[-0.200pt]{19.513pt}{0.400pt}}
\put(989.0,123.0){\rule[-0.200pt]{0.400pt}{3.132pt}}}
\put(181.0,123.0){\rule[-0.200pt]{194.647pt}{0.400pt}}
\put(989.0,123.0){\rule[-0.200pt]{0.400pt}{112.500pt}}
\put(181.0,590.0){\rule[-0.200pt]{194.647pt}{0.400pt}}
\put(181.0,123.0){\rule[-0.200pt]{0.400pt}{112.500pt}}
\end{picture}
\caption{Jet mass distributions for jets with $p_T>350$ GeV, for events that pass the cuts described in the text. 
We take  100 fb$^{-1}$ for the integrated luminosity.  }
\label{fig:jm1}
\end{figure}

In table \ref{table:second} we list the numbers of signal and 
background events that pass the successive cuts, again scaled to 100 
fb$^{-1}$.
The lepton cuts are especially effective in reducing the $t {\overline 
t}$ background, because when the highly boosted tops decay leptonically, 
the lepton is often too close to the $b$ quark to satisfy the isolation 
criteria.
\begin{table*}[htbp]
\begin{tabular}{||c||c|c|c|c||}
\hline
\hline
 &$N_{gen}$ & {$\sigma$(fb)} & {${\mathcal L}$(fb$^{-1}$)}&$N_{pass}$\\
\hline
  $t {\overline t} $+3 jets (inc.)  &  45,963 &  516 &89.0 & 80.8 \\
    $t {\overline t} $+2 jets  &  11,333 &  174 & 65.3 & 13.8 \\
      $t {\overline t} $+1 jets  &  5,686 &  $86.4 $ & 65.8 & 1.5\\
        $t {\overline t} $+0 jets  &  1,852 &  $34.0 $ & 98.1  & 0\\
 \hline
 $W$+4 jets (inc.) &  61,577 & 725 & 84.9  & 102\\
  $W$+3 jets  &  33,765 & 375 & 90.0  & 27.8\\
    $W$+2 jets  &  22,279 & 237 & 94.2  & 19.1\\
      $W$+1 jet  &  9,081 &  46.1  & 197 & 0\\
        $W$+0 jets  &  1,348 &  5.37 & 251 & 0\\
         \hline
 $Z$+4 jets (inc.) &  2,994 & $75.8 $ \ & 39.5 & 0 \\
  $Z$+3 jets  &  6,126 &  $40.6 $& 151  & 0.66\\
    $Z$+2 jets  &  3,716 &  $27.0 $ & 137 & 0 \\
      $Z$+1 jet  &  2,550 &  $5.20 $ & 490 & 0\\
                 \hline
  $WW$+3 jets (inc.)  &  9,471 & $105 $ & 90.1 & 20.0\\
    $WW$+2 jets  &  3,402 &   $42.4 $ & 80.3  & 8.7\\
      $WW$+1 jet  &  1,145 &  $13.2 $ & 86.5 & 0\\
        $WW$+0 jets  &  1,090&  $3.08 $& 354 & 0\\
\hline
  $ZZ$+3 jets (inc.)  &  212&   $1.93 $ & 110 & 0 \\
    $ZZ$+2 jets  &  77&   $0.679 $& 113  & 0\\
      $ZZ$+1 jet  &  55 &  $0.262 $ & 210 & 0\\
        $ZZ$+0 jets  &  125 &  $0.120 $& 1,041 & 0\\
\hline
  $W(l\nu)Z(f{\overline f})$+3 jets (inc.)  & 6,668 &   $26.3 $ & 253  & 9.1\\
    $W(l\nu)Z(f{\overline f})$+2 jets  &  3,703 &  $11.6 $ & 321  & 2.5\\
      $W(l\nu)Z(f{\overline f})$+1 jet  &  494 &  $3.50$ & 141 & 2.1\\
        $W(l\nu)Z(f{\overline f})$+0 jets  &  397 &  $0.421$& 942 & 0\\
\hline
  $W(q \overline{q})Z(l^+l^-)$+3 jets (inc.)  &  1,355 &   $5.45 $ & 249  & 0\\
    $W(q \overline{q})Z(l^+l^-)$+2 jets  &  354&   $2.24 $ & 158  & 0.63\\
      $W(q \overline{q})Z(l^+l^-)$+1 jet  &  299 &  $0.762 $ & 393 & 0\\
        $W(q \overline{q})Z(l^+l^-)$+0 jets  &  3,676 &  $8.12\times 10^{-2}$&$4.5 \times 10^4$  & 0\\
\hline
  $Wb{\overline b} $+2 jets (inc.)&  27,505&  177 & 155  & 102\\
    $Wb{\overline b} $+1 jet&  718 &12.0  & 59.8  & 13.4 \\
      $Wb{\overline b} $+0 jets&  446 & 1.53 & 291   & 1.4\\
\hline
    $tW$+1 jet   &  20,000 &  $335$ & 59.7  & 36.8\\
        $tW $  &  17,771 &  $78.1 $ & 228 & 12.3\\
      $tq $  &  5,487 &  $51.9 $ & 106  & 0\\
         $tb$  &  950 &  $9.72 $ & 97.7   & 0\\
        $tbW $ &  2,400 &  $60.6 $ & 39.6   & 15.2\\
        
\hline
  $B {\overline B} $ &  50,479 &$60.0 $& 841  & 210 \\
    $T {\overline T} $ ($HZ$) &  7,951 & $30.0 $ & 265 & 19.2 \\
      $T {\overline T} $   ($ZZ$)&  7,954 &  $14.1 $ & 564   & 14.9\\
        $T {\overline T} $  ($HH$)&  7,969 &  $16.0 $ & 498   & 6.8\\
\hline
\hline
\end{tabular}
\caption{For signal and background processes, the numbers of events generated $N_{gen}$, 
cross section, corresponding integrated luminosity, and number of events that pass all cuts $N_{pass}$,
rescaled to an integrated luminosity of 100 fb$^{-1}$.  For the background, the 
generator-level cuts described in the text are imposed.  For the 
samples labeled ``inc.'', extra jets from showering are allowed 
when the MLM prescription for jet-parton matching is applied. The cross 
sections in this table include $K$ factors to approximate NLO effects,
as described in text.}
\label{table:first}
\end{table*}
\begingroup
\squeezetable
\begin{table}[htbp]
\begin{tabular}{||c||c|c|c|c||}
\hline
\hline
 & signal(B) & $t {\overline t}$+ jets & $W$+ jets & $W b {\overline b}$+ jets \\
 \hline
generated &  6,000 & 80,995 & 138,801  & 19,053\\
 $\sum p_T > 1800\;$ GeV &  2,610 &  21,272 & 44,175 & 6,197  \\
lepton $p_T >100\;$ GeV&  864&  2,791 & 12,634  & 1,548 \\
${p_T \! \! \! \! \! \!\slash} \;> 100 \;$ GeV &  745& 2,035 &8,857 & 1014 \\
at least one $b$-tag & 387 &  1,009 & 483  & 302  \\
$\Delta R_{lj}>1.0$  &  246 &  182 & 314  & 210\\
$S_T>0.1$ &210 & 96 &149  & 117 \\
\hline
\hline
\end{tabular}
\caption{For the $B$ portion of the signal and the dominant background processes,
the numbers of events that pass the successive cuts, scaled to an integrated luminosity
of $100\;{\rm fb}^{-1}$.  For the background processes the first row gives the number
of events after the generator-level cut described  in the text.}
\label{table:second}
\end{table}
\endgroup
Jet mass distributions for the the signal and  two of the most important background processes,
$W$+jets and $t {\overline t}$+jets, are shown in figure \ref{fig:jm1}.   Only jets with $p_T>350$ GeV are 
included, and distributions are normalized to 100 fb$^{-1}$ of integrated luminosity.   
 The $W$ + jets distribution decreases significantly through the $W$ mass, the
$t {\overline t}$ + jets distribution is relatively flat in this region, 
while the signal has a pronounced peak around the $W$ mass.  
There is also a  smaller bump around the top mass, although the $t {\overline t}$ + jets distribution
also has a bump there due to the highly boosted tops.  The presence of highly boosted $W$'s 
also affects the   $t {\overline t}$+ jets distribution, but not nearly as dramatically as it does the signal.   
The distribution for $Wb{\overline b}$+ jets, the other large background,  steadily decreases as the jet mass increases.

\begin{figure}[htbp] 
  \centering
\setlength{\unitlength}{0.240900pt}
\ifx\plotpoint\undefined\newsavebox{\plotpoint}\fi
\begin{picture}(1050,630)(0,0)
\sbox{\plotpoint}{\rule[-0.200pt]{0.400pt}{0.400pt}}%
\put(181.0,123.0){\rule[-0.200pt]{4.818pt}{0.400pt}}
\put(161,123){\makebox(0,0)[r]{ 0}}
\put(969.0,123.0){\rule[-0.200pt]{4.818pt}{0.400pt}}
\put(181.0,190.0){\rule[-0.200pt]{4.818pt}{0.400pt}}
\put(161,190){\makebox(0,0)[r]{ 20}}
\put(969.0,190.0){\rule[-0.200pt]{4.818pt}{0.400pt}}
\put(181.0,256.0){\rule[-0.200pt]{4.818pt}{0.400pt}}
\put(161,256){\makebox(0,0)[r]{ 40}}
\put(969.0,256.0){\rule[-0.200pt]{4.818pt}{0.400pt}}
\put(181.0,323.0){\rule[-0.200pt]{4.818pt}{0.400pt}}
\put(161,323){\makebox(0,0)[r]{ 60}}
\put(969.0,323.0){\rule[-0.200pt]{4.818pt}{0.400pt}}
\put(181.0,390.0){\rule[-0.200pt]{4.818pt}{0.400pt}}
\put(161,390){\makebox(0,0)[r]{ 80}}
\put(969.0,390.0){\rule[-0.200pt]{4.818pt}{0.400pt}}
\put(181.0,457.0){\rule[-0.200pt]{4.818pt}{0.400pt}}
\put(161,457){\makebox(0,0)[r]{ 100}}
\put(969.0,457.0){\rule[-0.200pt]{4.818pt}{0.400pt}}
\put(181.0,523.0){\rule[-0.200pt]{4.818pt}{0.400pt}}
\put(161,523){\makebox(0,0)[r]{ 120}}
\put(969.0,523.0){\rule[-0.200pt]{4.818pt}{0.400pt}}
\put(181.0,590.0){\rule[-0.200pt]{4.818pt}{0.400pt}}
\put(161,590){\makebox(0,0)[r]{ 140}}
\put(969.0,590.0){\rule[-0.200pt]{4.818pt}{0.400pt}}
\put(181.0,123.0){\rule[-0.200pt]{0.400pt}{4.818pt}}
\put(181,82){\makebox(0,0){ 0}}
\put(181.0,570.0){\rule[-0.200pt]{0.400pt}{4.818pt}}
\put(316.0,123.0){\rule[-0.200pt]{0.400pt}{4.818pt}}
\put(316,82){\makebox(0,0){ 50}}
\put(316.0,570.0){\rule[-0.200pt]{0.400pt}{4.818pt}}
\put(450.0,123.0){\rule[-0.200pt]{0.400pt}{4.818pt}}
\put(450,82){\makebox(0,0){ 100}}
\put(450.0,570.0){\rule[-0.200pt]{0.400pt}{4.818pt}}
\put(585.0,123.0){\rule[-0.200pt]{0.400pt}{4.818pt}}
\put(585,82){\makebox(0,0){ 150}}
\put(585.0,570.0){\rule[-0.200pt]{0.400pt}{4.818pt}}
\put(720.0,123.0){\rule[-0.200pt]{0.400pt}{4.818pt}}
\put(720,82){\makebox(0,0){ 200}}
\put(720.0,570.0){\rule[-0.200pt]{0.400pt}{4.818pt}}
\put(854.0,123.0){\rule[-0.200pt]{0.400pt}{4.818pt}}
\put(854,82){\makebox(0,0){ 250}}
\put(854.0,570.0){\rule[-0.200pt]{0.400pt}{4.818pt}}
\put(989.0,123.0){\rule[-0.200pt]{0.400pt}{4.818pt}}
\put(989,82){\makebox(0,0){ 300}}
\put(989.0,570.0){\rule[-0.200pt]{0.400pt}{4.818pt}}
\put(181.0,123.0){\rule[-0.200pt]{194.647pt}{0.400pt}}
\put(989.0,123.0){\rule[-0.200pt]{0.400pt}{112.500pt}}
\put(181.0,590.0){\rule[-0.200pt]{194.647pt}{0.400pt}}
\put(181.0,123.0){\rule[-0.200pt]{0.400pt}{112.500pt}}
\put(60,356){\makebox(0,0){{\rotatebox{90}{Jets/Bin}}}}
\put(585,30){\makebox(0,0){Jet Mass (GeV)}}
\sbox{\plotpoint}{\rule[-0.400pt]{0.800pt}{0.800pt}}%
\sbox{\plotpoint}{\rule[-0.200pt]{0.400pt}{0.400pt}}%
\put(829,550){\makebox(0,0)[r]{{\tiny signal + background}}}
{\color{blue}\sbox{\plotpoint}{\rule[-0.400pt]{0.800pt}{0.800pt}}%
\put(849.0,550.0){\rule[-0.400pt]{24.090pt}{0.800pt}}
\put(181.0,123.0){\rule[-0.400pt]{0.800pt}{26.981pt}}
\put(181.0,235.0){\rule[-0.400pt]{6.504pt}{0.800pt}}
\put(208.0,235.0){\rule[-0.400pt]{0.800pt}{47.216pt}}
\put(208.0,431.0){\rule[-0.400pt]{6.504pt}{0.800pt}}
\put(235.0,431.0){\rule[-0.400pt]{0.800pt}{13.972pt}}
\put(235.0,489.0){\rule[-0.400pt]{6.504pt}{0.800pt}}
\put(262.0,423.0){\rule[-0.400pt]{0.800pt}{15.899pt}}
\put(262.0,423.0){\rule[-0.400pt]{6.504pt}{0.800pt}}
\put(289.0,343.0){\rule[-0.400pt]{0.800pt}{19.272pt}}
\put(289.0,343.0){\rule[-0.400pt]{6.504pt}{0.800pt}}
\put(316.0,331.0){\rule[-0.400pt]{0.800pt}{2.891pt}}
\put(316.0,331.0){\rule[-0.400pt]{6.504pt}{0.800pt}}
\put(343.0,331.0){\rule[-0.400pt]{0.800pt}{7.227pt}}
\put(343.0,361.0){\rule[-0.400pt]{6.504pt}{0.800pt}}
\put(370.0,361.0){\rule[-0.400pt]{0.800pt}{39.267pt}}
\put(370.0,524.0){\rule[-0.400pt]{6.263pt}{0.800pt}}
\put(396.0,423.0){\rule[-0.400pt]{0.800pt}{24.331pt}}
\put(396.0,423.0){\rule[-0.400pt]{6.504pt}{0.800pt}}
\put(423.0,356.0){\rule[-0.400pt]{0.800pt}{16.140pt}}
\put(423.0,356.0){\rule[-0.400pt]{6.504pt}{0.800pt}}
\put(450.0,288.0){\rule[-0.400pt]{0.800pt}{16.381pt}}
\put(450.0,288.0){\rule[-0.400pt]{6.504pt}{0.800pt}}
\put(477.0,269.0){\rule[-0.400pt]{0.800pt}{4.577pt}}
\put(477.0,269.0){\rule[-0.400pt]{6.504pt}{0.800pt}}
\put(504.0,238.0){\rule[-0.400pt]{0.800pt}{7.468pt}}
\put(504.0,238.0){\rule[-0.400pt]{6.504pt}{0.800pt}}
\put(531.0,238.0){\rule[-0.400pt]{0.800pt}{1.927pt}}
\put(531.0,246.0){\rule[-0.400pt]{6.504pt}{0.800pt}}
\put(558.0,246.0){\rule[-0.400pt]{0.800pt}{2.650pt}}
\put(558.0,257.0){\rule[-0.400pt]{6.504pt}{0.800pt}}
\put(585.0,255.0){\usebox{\plotpoint}}
\put(585.0,255.0){\rule[-0.400pt]{6.504pt}{0.800pt}}
\put(612.0,255.0){\rule[-0.400pt]{0.800pt}{13.009pt}}
\put(612.0,309.0){\rule[-0.400pt]{6.504pt}{0.800pt}}
\put(639.0,245.0){\rule[-0.400pt]{0.800pt}{15.418pt}}
\put(639.0,245.0){\rule[-0.400pt]{6.504pt}{0.800pt}}
\put(666.0,179.0){\rule[-0.400pt]{0.800pt}{15.899pt}}
\put(666.0,179.0){\rule[-0.400pt]{6.504pt}{0.800pt}}
\put(693.0,175.0){\usebox{\plotpoint}}
\put(693.0,175.0){\rule[-0.400pt]{6.504pt}{0.800pt}}
\put(720.0,168.0){\rule[-0.400pt]{0.800pt}{1.686pt}}
\put(720.0,168.0){\rule[-0.400pt]{6.504pt}{0.800pt}}
\put(747.0,168.0){\rule[-0.400pt]{0.800pt}{3.854pt}}
\put(747.0,184.0){\rule[-0.400pt]{6.504pt}{0.800pt}}
\put(774.0,154.0){\rule[-0.400pt]{0.800pt}{7.227pt}}
\put(774.0,154.0){\rule[-0.400pt]{6.263pt}{0.800pt}}
\put(800.0,145.0){\rule[-0.400pt]{0.800pt}{2.168pt}}
\put(800.0,145.0){\rule[-0.400pt]{6.504pt}{0.800pt}}
\put(827.0,145.0){\rule[-0.400pt]{0.800pt}{5.059pt}}
\put(827.0,166.0){\rule[-0.400pt]{6.504pt}{0.800pt}}
\put(854.0,141.0){\rule[-0.400pt]{0.800pt}{6.022pt}}
\put(854.0,141.0){\rule[-0.400pt]{6.504pt}{0.800pt}}
\put(881.0,141.0){\rule[-0.400pt]{0.800pt}{1.686pt}}
\put(881.0,148.0){\rule[-0.400pt]{6.504pt}{0.800pt}}
\put(908.0,140.0){\rule[-0.400pt]{0.800pt}{1.927pt}}
\put(908.0,140.0){\rule[-0.400pt]{6.504pt}{0.800pt}}
\put(935.0,140.0){\usebox{\plotpoint}}
\put(935.0,142.0){\rule[-0.400pt]{6.504pt}{0.800pt}}
\put(962.0,142.0){\usebox{\plotpoint}}
\put(962.0,146.0){\rule[-0.400pt]{6.504pt}{0.800pt}}
\put(989.0,123.0){\rule[-0.400pt]{0.800pt}{5.541pt}}}
\sbox{\plotpoint}{\rule[-0.200pt]{0.400pt}{0.400pt}}%
\put(829,509){\makebox(0,0)[r]{{\tiny background}}}
{\color{red}\put(849.0,509.0){\rule[-0.200pt]{24.090pt}{0.400pt}}
\put(181.0,123.0){\rule[-0.200pt]{0.400pt}{24.572pt}}
\put(181.0,225.0){\rule[-0.200pt]{6.504pt}{0.400pt}}
\put(208.0,225.0){\rule[-0.200pt]{0.400pt}{41.917pt}}
\put(208.0,399.0){\rule[-0.200pt]{6.504pt}{0.400pt}}
\put(235.0,399.0){\rule[-0.200pt]{0.400pt}{15.177pt}}
\put(235.0,462.0){\rule[-0.200pt]{6.504pt}{0.400pt}}
\put(262.0,395.0){\rule[-0.200pt]{0.400pt}{16.140pt}}
\put(262.0,395.0){\rule[-0.200pt]{6.504pt}{0.400pt}}
\put(289.0,313.0){\rule[-0.200pt]{0.400pt}{19.754pt}}
\put(289.0,313.0){\rule[-0.200pt]{6.504pt}{0.400pt}}
\put(316.0,284.0){\rule[-0.200pt]{0.400pt}{6.986pt}}
\put(316.0,284.0){\rule[-0.200pt]{6.504pt}{0.400pt}}
\put(343.0,267.0){\rule[-0.200pt]{0.400pt}{4.095pt}}
\put(343.0,267.0){\rule[-0.200pt]{6.504pt}{0.400pt}}
\put(370.0,267.0){\rule[-0.200pt]{0.400pt}{6.263pt}}
\put(370.0,293.0){\rule[-0.200pt]{6.263pt}{0.400pt}}
\put(396.0,227.0){\rule[-0.200pt]{0.400pt}{15.899pt}}
\put(396.0,227.0){\rule[-0.200pt]{6.504pt}{0.400pt}}
\put(423.0,227.0){\rule[-0.200pt]{0.400pt}{2.891pt}}
\put(423.0,239.0){\rule[-0.200pt]{6.504pt}{0.400pt}}
\put(450.0,220.0){\rule[-0.200pt]{0.400pt}{4.577pt}}
\put(450.0,220.0){\rule[-0.200pt]{6.504pt}{0.400pt}}
\put(477.0,211.0){\rule[-0.200pt]{0.400pt}{2.168pt}}
\put(477.0,211.0){\rule[-0.200pt]{6.504pt}{0.400pt}}
\put(504.0,192.0){\rule[-0.200pt]{0.400pt}{4.577pt}}
\put(504.0,192.0){\rule[-0.200pt]{6.504pt}{0.400pt}}
\put(531.0,191.0){\usebox{\plotpoint}}
\put(531.0,191.0){\rule[-0.200pt]{6.504pt}{0.400pt}}
\put(558.0,191.0){\rule[-0.200pt]{0.400pt}{1.445pt}}
\put(558.0,197.0){\rule[-0.200pt]{6.504pt}{0.400pt}}
\put(585.0,180.0){\rule[-0.200pt]{0.400pt}{4.095pt}}
\put(585.0,180.0){\rule[-0.200pt]{6.504pt}{0.400pt}}
\put(612.0,180.0){\rule[-0.200pt]{0.400pt}{10.400pt}}
\put(612.0,224.0){\rule[-0.200pt]{6.504pt}{0.400pt}}
\put(639.0,191.0){\rule[-0.200pt]{0.400pt}{7.950pt}}
\put(639.0,191.0){\rule[-0.200pt]{6.504pt}{0.400pt}}
\put(666.0,148.0){\rule[-0.200pt]{0.400pt}{10.359pt}}
\put(666.0,148.0){\rule[-0.200pt]{6.504pt}{0.400pt}}
\put(693.0,148.0){\rule[-0.200pt]{0.400pt}{1.445pt}}
\put(693.0,154.0){\rule[-0.200pt]{6.504pt}{0.400pt}}
\put(720.0,152.0){\rule[-0.200pt]{0.400pt}{0.482pt}}
\put(720.0,152.0){\rule[-0.200pt]{6.504pt}{0.400pt}}
\put(747.0,152.0){\rule[-0.200pt]{0.400pt}{4.336pt}}
\put(747.0,170.0){\rule[-0.200pt]{6.504pt}{0.400pt}}
\put(774.0,139.0){\rule[-0.200pt]{0.400pt}{7.468pt}}
\put(774.0,139.0){\rule[-0.200pt]{6.263pt}{0.400pt}}
\put(800.0,138.0){\usebox{\plotpoint}}
\put(800.0,138.0){\rule[-0.200pt]{6.504pt}{0.400pt}}
\put(827.0,138.0){\rule[-0.200pt]{0.400pt}{5.059pt}}
\put(827.0,159.0){\rule[-0.200pt]{6.504pt}{0.400pt}}
\put(854.0,133.0){\rule[-0.200pt]{0.400pt}{6.263pt}}
\put(854.0,133.0){\rule[-0.200pt]{6.504pt}{0.400pt}}
\put(881.0,133.0){\rule[-0.200pt]{0.400pt}{1.204pt}}
\put(881.0,138.0){\rule[-0.200pt]{6.504pt}{0.400pt}}
\put(908.0,134.0){\rule[-0.200pt]{0.400pt}{0.964pt}}
\put(908.0,134.0){\rule[-0.200pt]{6.504pt}{0.400pt}}
\put(935.0,134.0){\rule[-0.200pt]{0.400pt}{0.482pt}}
\put(935.0,136.0){\rule[-0.200pt]{6.504pt}{0.400pt}}
\put(962.0,136.0){\rule[-0.200pt]{0.400pt}{1.927pt}}
\put(962.0,144.0){\rule[-0.200pt]{6.504pt}{0.400pt}}
\put(989.0,123.0){\rule[-0.200pt]{0.400pt}{5.059pt}}}
\put(181.0,123.0){\rule[-0.200pt]{194.647pt}{0.400pt}}
\put(989.0,123.0){\rule[-0.200pt]{0.400pt}{112.500pt}}
\put(181.0,590.0){\rule[-0.200pt]{194.647pt}{0.400pt}}
\put(181.0,123.0){\rule[-0.200pt]{0.400pt}{112.500pt}}
\end{picture}
\caption{Jet mass distributions for the signal plus total background and for total background alone, 
for events that pass the cuts described in the text.  As before, only jets having $p_T>350$ GeV 
are included for each qualifying event, and we take  100 fb$^{-1}$ for the integrated luminosity. }
\label{fig:jm2}
\end{figure}
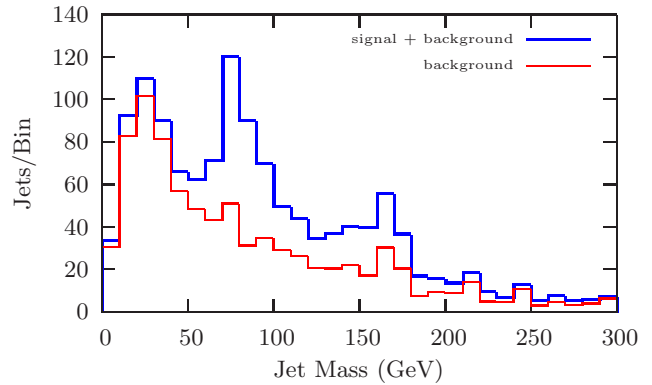

A plot of the  jet distributions for signal plus total background and total background alone is shown 
in figure \ref{fig:jm2}, again including only jets with $p_T>350$ GeV.
To estimate the significance of the peak around the $W$ mass, we take three times
the total number of jets in the 50--60 GeV bin (187) as a background value to compare
with the total number of jets in the 60--90 GeV bins (281), giving a 6.9$\sigma$ excess.  
More conservatively, taking the total number of jets in the  30--60 GeV bin (218) 
as the background value gives a 4.3$\sigma$ excess.  Finally, taking the total number 
of jets in the  40--70 GeV bin (200) as the background value for  the total number of jets in the 
70--100 GeV bins (280) gives a 5.7$\sigma$ excess.  For each of these three measures, 
the standard model contribution to the number of events in the peak is smaller than 
the standard model contribution to the estimated background value.  

The PGS detector simulator does not include particle deflection by the magnetic field, 
but to get a rough idea of how sensitive our results are to this effect, we
follow \cite{atlfast} and impose a shift in azimuthal angle for charged particles in the signal samples,
\begin{equation}
|\delta \phi| = \sin^{-1}(0.45/p_T),
\end{equation}
where the sign of the shift depends on the charge of the particle. We find that our results
are not dramatically affected by this shift.  The significance estimates above change 
to 6.8$\sigma$, 4.2$\sigma$, and 5.9$\sigma$, respectively.
 
The $T$ quarks do contribute somewhat to the signal, because their decays can produce $Z$ bosons,
which are not resolved from $W$'s using this method.  However, this contribution is relatively small. 
Recalculating the significance in each of the three ways described  previously, this time including only the $B$ contribution
to the signal, we find excesses  of  6.3$\sigma$,  3.5$\sigma$, and 4.5$\sigma$, respectively. 
 
We have seen that the jet mass distribution for the signal is peaked around the $W$ mass 
and less so around the top mass, due to the presence of highly boosted $W$'s and tops.  
Because the $B$ quark decays as $B\rightarrow t W^-$ (and the $T$ quark decays as $T\rightarrow t Z$ half of the time), 
one might hope to observe a peak in the invariant mass distribution of pairs of jets whose 
masses are near $m_W$ and $m_t$, respectively.   
\begin{figure}[htbp] 
  \centering
\setlength{\unitlength}{0.240900pt}
\ifx\plotpoint\undefined\newsavebox{\plotpoint}\fi
\begin{picture}(1050,630)(0,0)
\sbox{\plotpoint}{\rule[-0.200pt]{0.400pt}{0.400pt}}%
\put(161.0,123.0){\rule[-0.200pt]{4.818pt}{0.400pt}}
\put(141,123){\makebox(0,0)[r]{ 0}}
\put(969.0,123.0){\rule[-0.200pt]{4.818pt}{0.400pt}}
\put(161.0,175.0){\rule[-0.200pt]{4.818pt}{0.400pt}}
\put(141,175){\makebox(0,0)[r]{ 2}}
\put(969.0,175.0){\rule[-0.200pt]{4.818pt}{0.400pt}}
\put(161.0,227.0){\rule[-0.200pt]{4.818pt}{0.400pt}}
\put(141,227){\makebox(0,0)[r]{ 4}}
\put(969.0,227.0){\rule[-0.200pt]{4.818pt}{0.400pt}}
\put(161.0,279.0){\rule[-0.200pt]{4.818pt}{0.400pt}}
\put(141,279){\makebox(0,0)[r]{ 6}}
\put(969.0,279.0){\rule[-0.200pt]{4.818pt}{0.400pt}}
\put(161.0,331.0){\rule[-0.200pt]{4.818pt}{0.400pt}}
\put(141,331){\makebox(0,0)[r]{ 8}}
\put(969.0,331.0){\rule[-0.200pt]{4.818pt}{0.400pt}}
\put(161.0,382.0){\rule[-0.200pt]{4.818pt}{0.400pt}}
\put(141,382){\makebox(0,0)[r]{ 10}}
\put(969.0,382.0){\rule[-0.200pt]{4.818pt}{0.400pt}}
\put(161.0,434.0){\rule[-0.200pt]{4.818pt}{0.400pt}}
\put(141,434){\makebox(0,0)[r]{ 12}}
\put(969.0,434.0){\rule[-0.200pt]{4.818pt}{0.400pt}}
\put(161.0,486.0){\rule[-0.200pt]{4.818pt}{0.400pt}}
\put(141,486){\makebox(0,0)[r]{ 14}}
\put(969.0,486.0){\rule[-0.200pt]{4.818pt}{0.400pt}}
\put(161.0,538.0){\rule[-0.200pt]{4.818pt}{0.400pt}}
\put(141,538){\makebox(0,0)[r]{ 16}}
\put(969.0,538.0){\rule[-0.200pt]{4.818pt}{0.400pt}}
\put(161.0,590.0){\rule[-0.200pt]{4.818pt}{0.400pt}}
\put(141,590){\makebox(0,0)[r]{ 18}}
\put(969.0,590.0){\rule[-0.200pt]{4.818pt}{0.400pt}}
\put(161.0,123.0){\rule[-0.200pt]{0.400pt}{4.818pt}}
\put(161,82){\makebox(0,0){ 0}}
\put(161.0,570.0){\rule[-0.200pt]{0.400pt}{4.818pt}}
\put(244.0,123.0){\rule[-0.200pt]{0.400pt}{4.818pt}}
\put(244,82){\makebox(0,0){ 200}}
\put(244.0,570.0){\rule[-0.200pt]{0.400pt}{4.818pt}}
\put(327.0,123.0){\rule[-0.200pt]{0.400pt}{4.818pt}}
\put(327,82){\makebox(0,0){ 400}}
\put(327.0,570.0){\rule[-0.200pt]{0.400pt}{4.818pt}}
\put(409.0,123.0){\rule[-0.200pt]{0.400pt}{4.818pt}}
\put(409,82){\makebox(0,0){ 600}}
\put(409.0,570.0){\rule[-0.200pt]{0.400pt}{4.818pt}}
\put(492.0,123.0){\rule[-0.200pt]{0.400pt}{4.818pt}}
\put(492,82){\makebox(0,0){ 800}}
\put(492.0,570.0){\rule[-0.200pt]{0.400pt}{4.818pt}}
\put(575.0,123.0){\rule[-0.200pt]{0.400pt}{4.818pt}}
\put(575,82){\makebox(0,0){ 1000}}
\put(575.0,570.0){\rule[-0.200pt]{0.400pt}{4.818pt}}
\put(658.0,123.0){\rule[-0.200pt]{0.400pt}{4.818pt}}
\put(658,82){\makebox(0,0){ 1200}}
\put(658.0,570.0){\rule[-0.200pt]{0.400pt}{4.818pt}}
\put(741.0,123.0){\rule[-0.200pt]{0.400pt}{4.818pt}}
\put(741,82){\makebox(0,0){ 1400}}
\put(741.0,570.0){\rule[-0.200pt]{0.400pt}{4.818pt}}
\put(823.0,123.0){\rule[-0.200pt]{0.400pt}{4.818pt}}
\put(823,82){\makebox(0,0){ 1600}}
\put(823.0,570.0){\rule[-0.200pt]{0.400pt}{4.818pt}}
\put(906.0,123.0){\rule[-0.200pt]{0.400pt}{4.818pt}}
\put(906,82){\makebox(0,0){ 1800}}
\put(906.0,570.0){\rule[-0.200pt]{0.400pt}{4.818pt}}
\put(989.0,123.0){\rule[-0.200pt]{0.400pt}{4.818pt}}
\put(989,82){\makebox(0,0){ 2000}}
\put(989.0,570.0){\rule[-0.200pt]{0.400pt}{4.818pt}}
\put(161.0,123.0){\rule[-0.200pt]{199.465pt}{0.400pt}}
\put(989.0,123.0){\rule[-0.200pt]{0.400pt}{112.500pt}}
\put(161.0,590.0){\rule[-0.200pt]{199.465pt}{0.400pt}}
\put(161.0,123.0){\rule[-0.200pt]{0.400pt}{112.500pt}}
\put(60,356){\makebox(0,0){{\rotatebox{90}{Pairs/Bin}}}}
\put(575,30){\makebox(0,0){Invariant Mass (GeV)}}
\sbox{\plotpoint}{\rule[-0.400pt]{0.800pt}{0.800pt}}%
\sbox{\plotpoint}{\rule[-0.200pt]{0.400pt}{0.400pt}}%
\put(829,550){\makebox(0,0)[r]{{\tiny signal + background}}}
{\color{blue} \sbox{\plotpoint}{\rule[-0.400pt]{0.800pt}{0.800pt}}%
\put(849.0,550.0){\rule[-0.400pt]{24.090pt}{0.800pt}}
\put(161,123){\usebox{\plotpoint}}
\put(161.0,123.0){\rule[-0.400pt]{29.872pt}{0.800pt}}
\put(285.0,123.0){\rule[-0.400pt]{0.800pt}{1.445pt}}
\put(285.0,129.0){\rule[-0.400pt]{10.118pt}{0.800pt}}
\put(327.0,129.0){\rule[-0.400pt]{0.800pt}{2.168pt}}
\put(327.0,138.0){\rule[-0.400pt]{9.877pt}{0.800pt}}
\put(368.0,138.0){\rule[-0.400pt]{0.800pt}{13.249pt}}
\put(368.0,193.0){\rule[-0.400pt]{9.877pt}{0.800pt}}
\put(409.0,192.0){\usebox{\plotpoint}}
\put(409.0,192.0){\rule[-0.400pt]{10.118pt}{0.800pt}}
\put(451.0,192.0){\rule[-0.400pt]{0.800pt}{4.818pt}}
\put(451.0,212.0){\rule[-0.400pt]{9.877pt}{0.800pt}}
\put(492.0,212.0){\rule[-0.400pt]{0.800pt}{27.703pt}}
\put(492.0,327.0){\rule[-0.400pt]{10.118pt}{0.800pt}}
\put(534.0,327.0){\rule[-0.400pt]{0.800pt}{59.020pt}}
\put(534.0,572.0){\rule[-0.400pt]{9.877pt}{0.800pt}}
\put(575.0,291.0){\rule[-0.400pt]{0.800pt}{67.693pt}}
\put(575.0,291.0){\rule[-0.400pt]{9.877pt}{0.800pt}}
\put(616.0,291.0){\rule[-0.400pt]{0.800pt}{7.950pt}}
\put(616.0,324.0){\rule[-0.400pt]{10.118pt}{0.800pt}}
\put(658.0,242.0){\rule[-0.400pt]{0.800pt}{19.754pt}}
\put(658.0,242.0){\rule[-0.400pt]{9.877pt}{0.800pt}}
\put(699.0,242.0){\rule[-0.400pt]{0.800pt}{13.490pt}}
\put(699.0,298.0){\rule[-0.400pt]{10.118pt}{0.800pt}}
\put(741.0,204.0){\rule[-0.400pt]{0.800pt}{22.645pt}}
\put(741.0,204.0){\rule[-0.400pt]{9.877pt}{0.800pt}}
\put(782.0,204.0){\rule[-0.400pt]{0.800pt}{6.263pt}}
\put(782.0,230.0){\rule[-0.400pt]{9.877pt}{0.800pt}}
\put(823.0,209.0){\rule[-0.400pt]{0.800pt}{5.059pt}}
\put(823.0,209.0){\rule[-0.400pt]{10.118pt}{0.800pt}}
\put(865.0,209.0){\usebox{\plotpoint}}
\put(865.0,210.0){\rule[-0.400pt]{9.877pt}{0.800pt}}
\put(906.0,192.0){\rule[-0.400pt]{0.800pt}{4.336pt}}
\put(906.0,192.0){\rule[-0.400pt]{10.118pt}{0.800pt}}
\put(948.0,174.0){\rule[-0.400pt]{0.800pt}{4.336pt}}
\put(948.0,174.0){\rule[-0.400pt]{9.877pt}{0.800pt}}
\put(989.0,123.0){\rule[-0.400pt]{0.800pt}{12.286pt}}}
\sbox{\plotpoint}{\rule[-0.500pt]{1.000pt}{1.000pt}}%
\sbox{\plotpoint}{\rule[-0.200pt]{0.400pt}{0.400pt}}%
\put(829,509){\makebox(0,0)[r]{{\tiny background}}}
{\color{red} \sbox{\plotpoint}{\rule[-0.500pt]{1.000pt}{1.000pt}}%
\multiput(849,509)(20.756,0.000){5}{\usebox{\plotpoint}}
\put(949,509){\usebox{\plotpoint}}
\put(161,123){\usebox{\plotpoint}}
\multiput(161,123)(20.756,0.000){2}{\usebox{\plotpoint}}
\multiput(202,123)(20.756,0.000){2}{\usebox{\plotpoint}}
\multiput(244,123)(20.756,0.000){2}{\usebox{\plotpoint}}
\multiput(285,123)(20.756,0.000){2}{\usebox{\plotpoint}}
\multiput(327,123)(20.756,0.000){2}{\usebox{\plotpoint}}
\multiput(368,123)(0.000,20.756){2}{\usebox{\plotpoint}}
\multiput(368,156)(20.756,0.000){2}{\usebox{\plotpoint}}
\put(409.00,146.42){\usebox{\plotpoint}}
\multiput(409,133)(20.756,0.000){2}{\usebox{\plotpoint}}
\put(451.00,126.16){\usebox{\plotpoint}}
\multiput(451,123)(20.756,0.000){2}{\usebox{\plotpoint}}
\put(492.00,141.11){\usebox{\plotpoint}}
\multiput(492,152)(20.756,0.000){2}{\usebox{\plotpoint}}
\put(534.00,142.62){\usebox{\plotpoint}}
\multiput(534,134)(20.756,0.000){2}{\usebox{\plotpoint}}
\put(575.00,146.64){\usebox{\plotpoint}}
\multiput(575,161)(20.756,0.000){2}{\usebox{\plotpoint}}
\put(616.00,167.91){\usebox{\plotpoint}}
\multiput(616,182)(20.756,0.000){2}{\usebox{\plotpoint}}
\multiput(658,182)(0.000,-20.756){3}{\usebox{\plotpoint}}
\multiput(658,123)(20.756,0.000){2}{\usebox{\plotpoint}}
\multiput(699,123)(0.000,20.756){4}{\usebox{\plotpoint}}
\multiput(699,200)(20.756,0.000){2}{\usebox{\plotpoint}}
\multiput(741,200)(0.000,-20.756){3}{\usebox{\plotpoint}}
\multiput(741,140)(20.756,0.000){2}{\usebox{\plotpoint}}
\multiput(782,140)(0.000,20.756){2}{\usebox{\plotpoint}}
\multiput(782,183)(20.756,0.000){2}{\usebox{\plotpoint}}
\multiput(823,169)(20.756,0.000){2}{\usebox{\plotpoint}}
\put(865.00,171.80){\usebox{\plotpoint}}
\multiput(865,181)(20.756,0.000){2}{\usebox{\plotpoint}}
\put(906.00,168.94){\usebox{\plotpoint}}
\multiput(906,154)(20.756,0.000){2}{\usebox{\plotpoint}}
\multiput(948,152)(20.756,0.000){2}{\usebox{\plotpoint}}
\multiput(989,152)(0.000,-20.756){2}{\usebox{\plotpoint}}
\put(989,123){\usebox{\plotpoint}}}
\sbox{\plotpoint}{\rule[-0.200pt]{0.400pt}{0.400pt}}%
\put(161.0,123.0){\rule[-0.200pt]{199.465pt}{0.400pt}}
\put(989.0,123.0){\rule[-0.200pt]{0.400pt}{112.500pt}}
\put(161.0,590.0){\rule[-0.200pt]{199.465pt}{0.400pt}}
\put(161.0,123.0){\rule[-0.200pt]{0.400pt}{112.500pt}}
\end{picture}
\caption{Invariant mass distribution for pairs of $W$ and top candidates, after 100 fb$^{-1}$ of integrated luminosity.}
\label{fig:fulldijetF}
\end{figure}
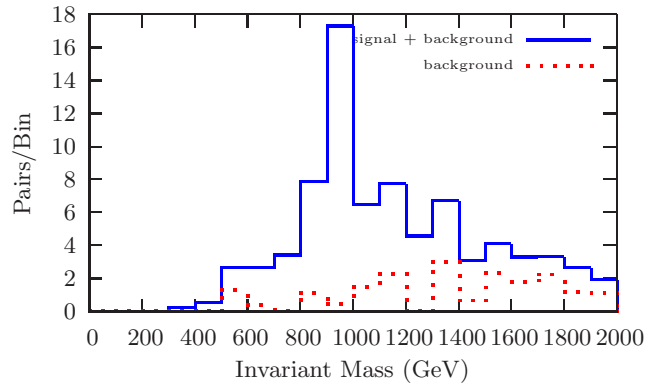
So, for each event
passing our cuts, we identify as $W$ candidates all jets with masses satisfying $|m_{jet} - m_W| < 20$ GeV, 
and we identify as top candidates all jets with masses satisfying $|m_{jet} - m_t| < 30$ GeV.  
Then, for each event we pair up the $W$ candidates with the top candidates in all possible ways, 
and calculate the invariant mass for each pairing.  A histogram of the resulting distribution is shown in figure \ref{fig:fulldijetF}. 

One simple point is that this procedure strongly enhances the ratio of signal to background.  
Beyond that, we see a clear peak near the heavy quark mass of 1 TeV.  This procedure tends
 to give a peak shifted somewhat below the actual mass.  After 100 fb$^{-1}$ of integrated luminosity,  
 fewer than $\sim 20$ pairs are obtained with invariant mass in the 900--1000 GeV bin, 
 but a more significant peak would be achieved with greater luminosity, or perhaps simply 
 by optimizing cuts and adjusting the jet mass windows used to identify candidate $W$'s and tops.  
 Alternatively, using jet mass in tandem with a more refined method for identifying tops 
 might enhance the signal. In our analysis there are fewer top candidates than there are $W$ 
 candidates, so to increase the significance one would first concentrate on enhancing the top signal.

\section{dilepton analysis}
The ratio of signal to background can be improved by  requiring two leptons 
(this also leaves essentially no hadronically decaying gauge bosons in the $t \overline t$ and $WW$  backgrounds).  
In this analysis, we impose the same cuts as before, except that now we require at least 
two leptons with $p_T > 20$ GeV and $|\eta|<2.5$, the hardest with $p_T > 100$ GeV.  
We will present results with and without the $b$-tag requirement.  
We do not consider the background source $W$+ jets with a fake second lepton.  

Without the $b$-tag requirement, 133 signal events remain (103 from $B$ production and decay), 
versus 104 background events.  The dominant backgrounds are $WW$+jets (40 events),  
$t {\overline t}+jets$ (28 events), $Z$+jets (16 events),   and $WZ$+jets (14 events).  
\begin{figure}[htbp] 
  \centering
\setlength{\unitlength}{0.240900pt}
\ifx\plotpoint\undefined\newsavebox{\plotpoint}\fi
\begin{picture}(1050,630)(0,0)
\sbox{\plotpoint}{\rule[-0.200pt]{0.400pt}{0.400pt}}%
\put(181.0,123.0){\rule[-0.200pt]{4.818pt}{0.400pt}}
\put(161,123){\makebox(0,0)[r]{ 0}}
\put(969.0,123.0){\rule[-0.200pt]{4.818pt}{0.400pt}}
\put(181.0,201.0){\rule[-0.200pt]{4.818pt}{0.400pt}}
\put(161,201){\makebox(0,0)[r]{ 20}}
\put(969.0,201.0){\rule[-0.200pt]{4.818pt}{0.400pt}}
\put(181.0,279.0){\rule[-0.200pt]{4.818pt}{0.400pt}}
\put(161,279){\makebox(0,0)[r]{ 40}}
\put(969.0,279.0){\rule[-0.200pt]{4.818pt}{0.400pt}}
\put(181.0,357.0){\rule[-0.200pt]{4.818pt}{0.400pt}}
\put(161,357){\makebox(0,0)[r]{ 60}}
\put(969.0,357.0){\rule[-0.200pt]{4.818pt}{0.400pt}}
\put(181.0,434.0){\rule[-0.200pt]{4.818pt}{0.400pt}}
\put(161,434){\makebox(0,0)[r]{ 80}}
\put(969.0,434.0){\rule[-0.200pt]{4.818pt}{0.400pt}}
\put(181.0,512.0){\rule[-0.200pt]{4.818pt}{0.400pt}}
\put(161,512){\makebox(0,0)[r]{ 100}}
\put(969.0,512.0){\rule[-0.200pt]{4.818pt}{0.400pt}}
\put(181.0,590.0){\rule[-0.200pt]{4.818pt}{0.400pt}}
\put(161,590){\makebox(0,0)[r]{ 120}}
\put(969.0,590.0){\rule[-0.200pt]{4.818pt}{0.400pt}}
\put(181.0,123.0){\rule[-0.200pt]{0.400pt}{4.818pt}}
\put(181,82){\makebox(0,0){ 0}}
\put(181.0,570.0){\rule[-0.200pt]{0.400pt}{4.818pt}}
\put(316.0,123.0){\rule[-0.200pt]{0.400pt}{4.818pt}}
\put(316,82){\makebox(0,0){ 50}}
\put(316.0,570.0){\rule[-0.200pt]{0.400pt}{4.818pt}}
\put(450.0,123.0){\rule[-0.200pt]{0.400pt}{4.818pt}}
\put(450,82){\makebox(0,0){ 100}}
\put(450.0,570.0){\rule[-0.200pt]{0.400pt}{4.818pt}}
\put(585.0,123.0){\rule[-0.200pt]{0.400pt}{4.818pt}}
\put(585,82){\makebox(0,0){ 150}}
\put(585.0,570.0){\rule[-0.200pt]{0.400pt}{4.818pt}}
\put(720.0,123.0){\rule[-0.200pt]{0.400pt}{4.818pt}}
\put(720,82){\makebox(0,0){ 200}}
\put(720.0,570.0){\rule[-0.200pt]{0.400pt}{4.818pt}}
\put(854.0,123.0){\rule[-0.200pt]{0.400pt}{4.818pt}}
\put(854,82){\makebox(0,0){ 250}}
\put(854.0,570.0){\rule[-0.200pt]{0.400pt}{4.818pt}}
\put(989.0,123.0){\rule[-0.200pt]{0.400pt}{4.818pt}}
\put(989,82){\makebox(0,0){ 300}}
\put(989.0,570.0){\rule[-0.200pt]{0.400pt}{4.818pt}}
\put(181.0,123.0){\rule[-0.200pt]{194.647pt}{0.400pt}}
\put(989.0,123.0){\rule[-0.200pt]{0.400pt}{112.500pt}}
\put(181.0,590.0){\rule[-0.200pt]{194.647pt}{0.400pt}}
\put(181.0,123.0){\rule[-0.200pt]{0.400pt}{112.500pt}}
\put(90,356){\makebox(0,0){\rotatebox{90}{Jets/Bin}}}
\put(585,30){\makebox(0,0){Jet Mass (GeV)}}
\sbox{\plotpoint}{\rule[-0.400pt]{0.800pt}{0.800pt}}%
\sbox{\plotpoint}{\rule[-0.200pt]{0.400pt}{0.400pt}}%
\put(829,550){\makebox(0,0)[r]{{\tiny signal + background}}}
\sbox{\plotpoint}{\rule[-0.400pt]{0.800pt}{0.800pt}}%
{\color{blue} \put(849.0,550.0){\rule[-0.400pt]{24.090pt}{0.800pt}}
\put(181.0,123.0){\rule[-0.400pt]{0.800pt}{49.144pt}}
\put(181.0,327.0){\rule[-0.400pt]{19.513pt}{0.800pt}}
\put(262.0,327.0){\rule[-0.400pt]{0.800pt}{10.600pt}}
\put(262.0,371.0){\rule[-0.400pt]{19.513pt}{0.800pt}}
\put(343.0,371.0){\rule[-0.400pt]{0.800pt}{37.580pt}}
\put(343.0,527.0){\rule[-0.400pt]{19.272pt}{0.800pt}}
\put(423.0,343.0){\rule[-0.400pt]{0.800pt}{44.326pt}}
\put(423.0,343.0){\rule[-0.400pt]{19.513pt}{0.800pt}}
\put(504.0,268.0){\rule[-0.400pt]{0.800pt}{18.067pt}}
\put(504.0,268.0){\rule[-0.400pt]{19.513pt}{0.800pt}}
\put(585.0,268.0){\rule[-0.400pt]{0.800pt}{7.468pt}}
\put(585.0,299.0){\rule[-0.400pt]{19.513pt}{0.800pt}}
\put(666.0,187.0){\rule[-0.400pt]{0.800pt}{26.981pt}}
\put(666.0,187.0){\rule[-0.400pt]{19.513pt}{0.800pt}}
\put(747.0,171.0){\rule[-0.400pt]{0.800pt}{3.854pt}}
\put(747.0,171.0){\rule[-0.400pt]{19.272pt}{0.800pt}}
\put(827.0,143.0){\rule[-0.400pt]{0.800pt}{6.745pt}}
\put(827.0,143.0){\rule[-0.400pt]{19.513pt}{0.800pt}}
\put(908.0,137.0){\rule[-0.400pt]{0.800pt}{1.445pt}}
\put(908.0,137.0){\rule[-0.400pt]{19.513pt}{0.800pt}}
\put(989.0,123.0){\rule[-0.400pt]{0.800pt}{3.373pt}}}
\sbox{\plotpoint}{\rule[-0.500pt]{1.000pt}{1.000pt}}%
\sbox{\plotpoint}{\rule[-0.200pt]{0.400pt}{0.400pt}}%
\put(829,509){\makebox(0,0)[r]{{\tiny background}}}
{\color{red} \sbox{\plotpoint}{\rule[-0.500pt]{1.000pt}{1.000pt}}%
 \multiput(849,509)(20.756,0.000){5}{\usebox{\plotpoint}}
\put(949,509){\usebox{\plotpoint}}
\multiput(181,123)(0.000,20.756){7}{\usebox{\plotpoint}}
\multiput(181,252)(20.756,0.000){4}{\usebox{\plotpoint}}
\multiput(262,252)(0.000,20.756){2}{\usebox{\plotpoint}}
\multiput(262,301)(20.756,0.000){4}{\usebox{\plotpoint}}
\multiput(343,301)(0.000,-20.756){3}{\usebox{\plotpoint}}
\multiput(343,229)(20.756,0.000){4}{\usebox{\plotpoint}}
\put(423.00,222.87){\usebox{\plotpoint}}
\multiput(423,204)(20.756,0.000){4}{\usebox{\plotpoint}}
\put(504.00,200.09){\usebox{\plotpoint}}
\multiput(504,189)(20.756,0.000){4}{\usebox{\plotpoint}}
\multiput(585,182)(20.756,0.000){4}{\usebox{\plotpoint}}
\multiput(666,182)(0.000,-20.756){2}{\usebox{\plotpoint}}
\multiput(666,143)(20.756,0.000){4}{\usebox{\plotpoint}}
\put(747.00,154.24){\usebox{\plotpoint}}
\multiput(747,155)(20.756,0.000){3}{\usebox{\plotpoint}}
\multiput(827,155)(0.000,-20.756){2}{\usebox{\plotpoint}}
\multiput(827,129)(20.756,0.000){4}{\usebox{\plotpoint}}
\multiput(908,129)(20.756,0.000){3}{\usebox{\plotpoint}}
\put(989.00,127.94){\usebox{\plotpoint}}
\put(989,123){\usebox{\plotpoint}}}
\sbox{\plotpoint}{\rule[-0.200pt]{0.400pt}{0.400pt}}%
\put(181.0,123.0){\rule[-0.200pt]{194.647pt}{0.400pt}}
\put(989.0,123.0){\rule[-0.200pt]{0.400pt}{112.500pt}}
\put(181.0,590.0){\rule[-0.200pt]{194.647pt}{0.400pt}}
\put(181.0,123.0){\rule[-0.200pt]{0.400pt}{112.500pt}}
\end{picture}
\caption{Jet mass distributions for the signal plus total background and for total background alone, 
for events passing the cuts in the dilepton analysis.  Only jets having $p_T>300$ GeV are included 
for each qualifying event, and we take  100 fb$^{-1}$ for the integrated luminosity.}
\label{fig:dilepF}
\end{figure}
The resulting jet-mass distributions are are shown in figure  \ref{fig:dilepF}, this time keeping only jets 
with $p_T>300$ GeV. Taking the total number of jets in the 30--60 GeV bin (66) as a background value for the
total number of jets in the 60--90 GeV bin (102), we find a  $4.4\sigma$ excess.  

\begin{figure}[htbp] 
  \centering
\setlength{\unitlength}{0.240900pt}
\ifx\plotpoint\undefined\newsavebox{\plotpoint}\fi
\begin{picture}(1050,630)(0,0)
\sbox{\plotpoint}{\rule[-0.200pt]{0.400pt}{0.400pt}}%
\put(161.0,123.0){\rule[-0.200pt]{4.818pt}{0.400pt}}
\put(141,123){\makebox(0,0)[r]{ 0}}
\put(969.0,123.0){\rule[-0.200pt]{4.818pt}{0.400pt}}
\put(161.0,170.0){\rule[-0.200pt]{4.818pt}{0.400pt}}
\put(141,170){\makebox(0,0)[r]{ 5}}
\put(969.0,170.0){\rule[-0.200pt]{4.818pt}{0.400pt}}
\put(161.0,216.0){\rule[-0.200pt]{4.818pt}{0.400pt}}
\put(141,216){\makebox(0,0)[r]{ 10}}
\put(969.0,216.0){\rule[-0.200pt]{4.818pt}{0.400pt}}
\put(161.0,263.0){\rule[-0.200pt]{4.818pt}{0.400pt}}
\put(141,263){\makebox(0,0)[r]{ 15}}
\put(969.0,263.0){\rule[-0.200pt]{4.818pt}{0.400pt}}
\put(161.0,310.0){\rule[-0.200pt]{4.818pt}{0.400pt}}
\put(141,310){\makebox(0,0)[r]{ 20}}
\put(969.0,310.0){\rule[-0.200pt]{4.818pt}{0.400pt}}
\put(161.0,357.0){\rule[-0.200pt]{4.818pt}{0.400pt}}
\put(141,357){\makebox(0,0)[r]{ 25}}
\put(969.0,357.0){\rule[-0.200pt]{4.818pt}{0.400pt}}
\put(161.0,403.0){\rule[-0.200pt]{4.818pt}{0.400pt}}
\put(141,403){\makebox(0,0)[r]{ 30}}
\put(969.0,403.0){\rule[-0.200pt]{4.818pt}{0.400pt}}
\put(161.0,450.0){\rule[-0.200pt]{4.818pt}{0.400pt}}
\put(141,450){\makebox(0,0)[r]{ 35}}
\put(969.0,450.0){\rule[-0.200pt]{4.818pt}{0.400pt}}
\put(161.0,497.0){\rule[-0.200pt]{4.818pt}{0.400pt}}
\put(141,497){\makebox(0,0)[r]{ 40}}
\put(969.0,497.0){\rule[-0.200pt]{4.818pt}{0.400pt}}
\put(161.0,543.0){\rule[-0.200pt]{4.818pt}{0.400pt}}
\put(141,543){\makebox(0,0)[r]{ 45}}
\put(969.0,543.0){\rule[-0.200pt]{4.818pt}{0.400pt}}
\put(161.0,590.0){\rule[-0.200pt]{4.818pt}{0.400pt}}
\put(141,590){\makebox(0,0)[r]{ 50}}
\put(969.0,590.0){\rule[-0.200pt]{4.818pt}{0.400pt}}
\put(161.0,123.0){\rule[-0.200pt]{0.400pt}{4.818pt}}
\put(161,82){\makebox(0,0){ 0}}
\put(161.0,570.0){\rule[-0.200pt]{0.400pt}{4.818pt}}
\put(299.0,123.0){\rule[-0.200pt]{0.400pt}{4.818pt}}
\put(299,82){\makebox(0,0){ 50}}
\put(299.0,570.0){\rule[-0.200pt]{0.400pt}{4.818pt}}
\put(437.0,123.0){\rule[-0.200pt]{0.400pt}{4.818pt}}
\put(437,82){\makebox(0,0){ 100}}
\put(437.0,570.0){\rule[-0.200pt]{0.400pt}{4.818pt}}
\put(575.0,123.0){\rule[-0.200pt]{0.400pt}{4.818pt}}
\put(575,82){\makebox(0,0){ 150}}
\put(575.0,570.0){\rule[-0.200pt]{0.400pt}{4.818pt}}
\put(713.0,123.0){\rule[-0.200pt]{0.400pt}{4.818pt}}
\put(713,82){\makebox(0,0){ 200}}
\put(713.0,570.0){\rule[-0.200pt]{0.400pt}{4.818pt}}
\put(851.0,123.0){\rule[-0.200pt]{0.400pt}{4.818pt}}
\put(851,82){\makebox(0,0){ 250}}
\put(851.0,570.0){\rule[-0.200pt]{0.400pt}{4.818pt}}
\put(989.0,123.0){\rule[-0.200pt]{0.400pt}{4.818pt}}
\put(989,82){\makebox(0,0){ 300}}
\put(989.0,570.0){\rule[-0.200pt]{0.400pt}{4.818pt}}
\put(161.0,123.0){\rule[-0.200pt]{199.465pt}{0.400pt}}
\put(989.0,123.0){\rule[-0.200pt]{0.400pt}{112.500pt}}
\put(161.0,590.0){\rule[-0.200pt]{199.465pt}{0.400pt}}
\put(161.0,123.0){\rule[-0.200pt]{0.400pt}{112.500pt}}
\put(60,356){\makebox(0,0){{\rotatebox{90}{Jets/Bin}}}}
\put(575,30){\makebox(0,0){Jet Mass (GeV)}}
\sbox{\plotpoint}{\rule[-0.400pt]{0.800pt}{0.800pt}}%
\sbox{\plotpoint}{\rule[-0.200pt]{0.400pt}{0.400pt}}%
\put(829,550){\makebox(0,0)[r]{{\tiny signal + background}}}
{\color{blue} \sbox{\plotpoint}{\rule[-0.400pt]{0.800pt}{0.800pt}}%
\put(849.0,550.0){\rule[-0.400pt]{24.090pt}{0.800pt}}
\put(161.0,123.0){\rule[-0.400pt]{0.800pt}{35.412pt}}
\put(161.0,270.0){\rule[-0.400pt]{19.995pt}{0.800pt}}
\put(244.0,256.0){\rule[-0.400pt]{0.800pt}{3.373pt}}
\put(244.0,256.0){\rule[-0.400pt]{19.995pt}{0.800pt}}
\put(327.0,256.0){\rule[-0.400pt]{0.800pt}{69.861pt}}
\put(327.0,546.0){\rule[-0.400pt]{19.754pt}{0.800pt}}
\put(409.0,309.0){\rule[-0.400pt]{0.800pt}{57.093pt}}
\put(409.0,309.0){\rule[-0.400pt]{19.995pt}{0.800pt}}
\put(492.0,266.0){\rule[-0.400pt]{0.800pt}{10.359pt}}
\put(492.0,266.0){\rule[-0.400pt]{19.995pt}{0.800pt}}
\put(575.0,266.0){\rule[-0.400pt]{0.800pt}{7.227pt}}
\put(575.0,296.0){\rule[-0.400pt]{19.995pt}{0.800pt}}
\put(658.0,196.0){\rule[-0.400pt]{0.800pt}{24.090pt}}
\put(658.0,196.0){\rule[-0.400pt]{19.995pt}{0.800pt}}
\put(741.0,146.0){\rule[-0.400pt]{0.800pt}{12.045pt}}
\put(741.0,146.0){\rule[-0.400pt]{19.754pt}{0.800pt}}
\put(823.0,143.0){\usebox{\plotpoint}}
\put(823.0,143.0){\rule[-0.400pt]{19.995pt}{0.800pt}}
\put(906.0,135.0){\rule[-0.400pt]{0.800pt}{1.927pt}}
\put(906.0,135.0){\rule[-0.400pt]{19.995pt}{0.800pt}}
\put(989.0,123.0){\rule[-0.400pt]{0.800pt}{2.891pt}}}
\sbox{\plotpoint}{\rule[-0.500pt]{1.000pt}{1.000pt}}%
\sbox{\plotpoint}{\rule[-0.200pt]{0.400pt}{0.400pt}}%
\put(829,509){\makebox(0,0)[r]{{\tiny background}}}
{\color{red} \sbox{\plotpoint}{\rule[-0.500pt]{1.000pt}{1.000pt}}%
\multiput(849,509)(20.756,0.000){5}{\usebox{\plotpoint}}
\put(949,509){\usebox{\plotpoint}}
\multiput(161,123)(0.000,20.756){3}{\usebox{\plotpoint}}
\multiput(161,174)(20.756,0.000){4}{\usebox{\plotpoint}}
\put(244.00,162.71){\usebox{\plotpoint}}
\multiput(244,159)(20.756,0.000){4}{\usebox{\plotpoint}}
\multiput(327,152)(20.756,0.000){4}{\usebox{\plotpoint}}
\multiput(409,144)(20.756,0.000){4}{\usebox{\plotpoint}}
\put(492.00,147.11){\usebox{\plotpoint}}
\multiput(492,158)(20.756,0.000){4}{\usebox{\plotpoint}}
\put(575.00,148.11){\usebox{\plotpoint}}
\multiput(575,142)(20.756,0.000){4}{\usebox{\plotpoint}}
\multiput(658,144)(20.756,0.000){4}{\usebox{\plotpoint}}
\put(741.00,131.31){\usebox{\plotpoint}}
\multiput(741,123)(20.756,0.000){4}{\usebox{\plotpoint}}
\multiput(823,123)(20.756,0.000){4}{\usebox{\plotpoint}}
\multiput(906,123)(20.756,0.000){4}{\usebox{\plotpoint}}
\put(989,123){\usebox{\plotpoint}}}
\sbox{\plotpoint}{\rule[-0.200pt]{0.400pt}{0.400pt}}%
\put(161.0,123.0){\rule[-0.200pt]{199.465pt}{0.400pt}}
\put(989.0,123.0){\rule[-0.200pt]{0.400pt}{112.500pt}}
\put(161.0,590.0){\rule[-0.200pt]{199.465pt}{0.400pt}}
\put(161.0,123.0){\rule[-0.200pt]{0.400pt}{112.500pt}}
\end{picture}
\caption{Same as figure \ref{fig:dilepF}, but with a $b$-tag requirement.}
\label{fig:dilepF2}
\end{figure}
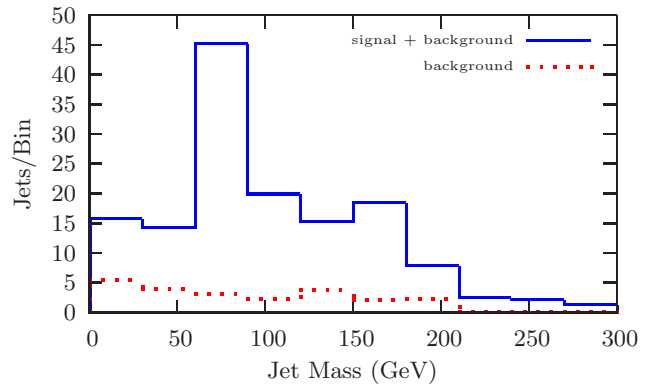

After a $b$-tag is required, 72 signal events remain (54 from $B$ production and decay), versus
only 12 background events. The jet mass distribution,  shown in figure  \ref{fig:dilepF2},  
has a  peak with greater than $5\sigma$ significance.
%
%

\section{conclusions}
Jet mass distributions will provide a useful probe of new physics at the LHC.
In particular, we have seen that they give signals in excess of 5$\sigma$ for vector-quark doublets of TeV mass
after 100 fb$^{-1}$ of integrated luminosity.  

It would be interesting to try to use jet mass to  test other models.
 There may be parameter space in supersymmetric models for which 
methods similar to the ones we have used would be effective, 
{\em e.g.} if  heavy charginos are produced copiously and decay dominantly 
to $W$ bosons and LSP neutralinos.  Another possible application is to warped-space models \cite{Randall:1999ee}.
For example, in ref. \cite{Dennis:2007tv} the detection of Kaluza-Klein bottom quarks, which decay in the same way
as the $B$ quarks considered here, was considered for masses in the $\sim 500$ GeV range. 
For heavier masses the methods outlined here would be useful. 
In ref. \cite{Agashe:2007zd}, the detection of Kaluza-Klein gravitons through their 
decays into gauge bosons was studied, and jet mass distributions  might be helpful there as well.  
Finally, techniques for dealing with highly-boosted  tops in the context of warped models 
have been proposed in \cite{Agashe:2006hk}.  It is possible that jet mass considerations
could also help for this purpose.  

\section*{Note added}
A week after we submitted this article to the hep-ph arxiv, Ref.~\cite{BHoldom}
appeared.  We would like to draw readers' attention to
this work, as it also uses jet mass as a probe of new physics.

\section*{Acknowledgements}
The work of WS was supported in part by the Department of Energy under Grant 
DE-FG-02-92ER40704.  The work of DTS was supported in part by NSF grant PHY-0555421.

\end{document}